\documentclass[twocolumn,preprintnumbers,amsmath,amssymb,showpacs]{revtex4}
\usepackage{graphicx}
\usepackage{dcolumn}
\usepackage{bm}
\usepackage{epsf}
\usepackage{amsmath}
\usepackage{amsfonts}

\newcommand{\DDir}{\relax{D\kern-.7em{/}}}

\newcommand{\soo}{\Rightarrow}

\newcommand{\be}{\begin{equation}}
\newcommand{\ee}{\end{equation}}
\newcommand{\bea}{\begin{equation*}}
\newcommand{\eea}{\end{equation*}}

\newcommand{\nin}{\relax{\in\kern-.8em{/}}}

\newcommand{\lm}{\lambda}

\newcommand{\om}{\omega}
\newcommand{\sig}{\sigma}

\newcommand{\vep}{\varepsilon}
\newcommand{\ep}{\epsilon}

\newcommand{\cm}{\mbox{ cm}}

\newcommand{\erg}{\mbox{ erg}}

\newcommand{\Mpc}{\mbox{ Mpc}}
\newcommand{\eV}{\mbox{eV}}
\newcommand{\keV}{\mbox{ keV}}
\newcommand{\MeV}{\mbox{ MeV}}

\newcommand{\TeV}{\mbox{ TeV}}

\newcommand{\dy}{\mbox{d}}

\newcommand{\Rbr}{R_{\rm br}}
\newcommand{\apjl}{ApJ}

\begin{document}
\title{X-rays, $\gamma$-rays and neutrinos from collisoinless shocks in supernova wind breakouts}
\author{Boaz Katz$^{1}$\footnote{Bahcal Fellow, Einstein Fellow}}
\author{Nir Sapir$^{2}$}
\author{Eli Waxman$^{2}$}
\affiliation{$^1$Institute for Advanced Study, Princeton, NJ 08540, USA}
\affiliation{$^2$Dept. of Particle Phys. \& Astrophys., Weizmann Institute of Science, Rehovot 76100, Israel}

\begin{abstract}
We show that a collisionless shock necessarily forms during the shock breakout of a supernova (SN) surrounded by an optically thick wind. An intense non-thermal flash of $\lesssim1 \MeV$ gamma rays, hard X-rays and multi-TeV neutrinos is produced simultaneously with and following the soft X-ray breakout emission, carrying similar or larger energy than the soft emission. The non-thermal flash is detectable by current X-ray telescopes and may be detectable out to 10's of Mpc by km-scale neutrino telescopes.\end{abstract}
\maketitle

It has long been suggested that an intense burst of  X-ray radiation is expected to be emitted at the initial phases of a SN explosion, once the radiation mediated blast wave reaches the edge of the star \cite{Colgate74,Falk78,Klein78,Epstein81,Ensman92,Katz10}. If the star is surrounded by a sufficiently optically thick shell of circum-stellar matter (CSM), e.g. a high density wind, the breakout occurs within the shell. Several observed $\gamma$-ray/X-ray flashes associated with SNe \cite{Campana06,Soderberg08} have been suggested \cite{Campana06,Waxman07,Wang07,Soderberg08,Katz10,Balberg11} to be such wind breakouts of fast, $v_{\rm sh}\gtrsim 0.1c$, shocks, in which departure from equilibrium may imply very high electron temperatures reaching tens or hundreds of $\keV$ \citep{Katz10,Nakar10}.
In fact, all low luminosity $\gamma$-ray bursts associated with SNe may be produced by such fast, $v_{\rm sh}\gtrsim 0.1c$, breakouts (with or without the presence of an optically thick CSM) \cite{Waxman07,Wang07,Katz10}. 
Breakout outbursts of slower shocks, $v_{\rm sh}\sim 0.03c$, have been suggested to account for strong optical/UV transients \cite{Ofek10} and very luminous SNe \cite[e.g.][]{Quimby07,Smith07a,Miller09}.
In order to explain the high energy (reaching $10^{51}\erg$) emitted in these SNe, CSM parameters were suggested such that the diffusion time scale through the CSM is comparable to the dynamical time scale, $R/v_{\rm sh}$ \cite [e.g.][]{Quimby07,Smith07b,Moria11,Chevalier11}. If true, the observed emission is, by construction, the breakout outburst from the CSM.

Following breakout, the radiation mediated shock is expected to become a collisionless shock, leading to the emission of gamma-rays and neutrinos \cite{Waxman01}. In the absence of a (significant) wind, the small mass of the shell shocked by the collisionless shock implies that only a small fraction, $\lesssim 10^{-2}$, of the breakout energy is converted to such high energy radiation. Moreover, the formation of a collisionless (or collisional) shock is controversial \cite[e.g.][]{Klein78,Lasher79,Epstein81,Blinnikov91,Sapir11,Ensman92}, since the light shell may be accelerated to sufficiently high velocity by the escaping radiation. In this letter we show that if the progenitor is surrounded by an optically thick CSM, e.g. a dense wind, a collsionless shock is necessarily created during the breakout, and that an energy comparable to or greater than the breakout energy is emitted by quasi-thermal particles in high energy ($\gtrsim 50\keV$) photons, and by accelerated protons in high energy ($\gtrsim1~\TeV$) neutrinos. The latter is an extension of the study of high energy emission from the interaction of the ejecta with a dense optically thin CSM \cite{Murase10}.

\paragraph*{Formation of a collisionless shock.}\label{sec:ColShockForm}
Consider first for simplicity a piston moving with a constant velocity $v=10^9\cm\sec^{-1}v_9$ through an optically thick fully ionized hydrogen wind with a density profile
\begin{equation}
\rho(r)=\frac{c}{v}\frac{m_p}{\sig_T\Rbr}(r/\Rbr)^{-2},
\end{equation}
where $\Rbr=10^{14}R_{14}\cm$ is a normalization parameter with dimensions of length.   A shock propagates ahead of the piston with velocity $v_{\rm sh}\sim v$.
As long as the optical depth across the shock transition region, $\Delta_{\tau,\rm sh}\sim c/v$, is much smaller than the optical depth of the system, $\Delta_{\tau}=(c/v)\Rbr/r$, the post-shock radiation is confined. Once the shock reaches $r\sim\Rbr$, the width of the shock becomes comparable to the size of the system and a significant fraction of the post-shock energy can be emitted during one dynamical time scale, $\Rbr/v$. This emission is the breakout outburst discussed above.

The material lying ahead of the piston must be accelerated to velocities approacing $v$ by some process. At large optical depth, where the radiation mediated shock is sustained, the radiation accelerates the material by Compton scattering off the electrons. The maximal velocity to which  a fluid shell can be accelerated by this process is given by
\begin{equation}\label{eq:vmax}
v_{\max}=\frac{E_{\gamma}/c}{4\pi r^2}\frac{\sig_T}{m_p},
\end{equation}
where $E_{\gamma}=\int L_{\gamma}dt$ is the radiation energy emitted through the fluid shell and $r$ is its initial position. This maximum velocity is achieved if all of the flowing photons move radially and the shell does not expand considerably during the passage of the radiation. In this case, a fraction $\sig_T/(4\pi r^2)$ of the momentum $E_{\gamma}/c$ carried by the radiation is transferred on average to each proton.

$E_{\gamma}$ is limited by the thermal energy accumulated in the post shock region which, in turn, is limited by $0.5 M(r) v^2$, where $M(r)=4\pi (c/\kappa v)\Rbr r$ is the wind mass inward of $r$. Using equation \eqref{eq:vmax} an upper limit for $v_{\max}$, $v_{\max}<0.5 (\Rbr/r)v$, is obtained. This implies that beyond a radius of $0.5\Rbr$ the shock can no longer be mediated by radiation and must be transformed into a collisional or a collisionless shock. Since the ion plasma frequency, $\om_{p}=(4\pi \rho e^2/m_p^2)^{1/2}\sim 10^{9}v_9^{-1/2}R_{14}^{-1/2}\sec^{-1}$, is many orders of magnitudes larger than the ion Coulomb collision rate per particle, $\nu_C=\rho\sig_Cv/m_p\sim 2\times 10^{-2}R_{14}^{-1}v_9^{-4}\sec^{-1}$ \citep[e.g.][]{Waxman01}, the shock will be collisionless, i.e. mediated by collective plasma instabilities. 

Let us comment on our simplifying assumptions of constant piston velocity and wind CSM profile. In reality, the shock velocity slowly changes with time as the faster parts of the SN ejecta are slowed down by the wind. The arguments above hold if the constant $v$ is replaced by its value at the vicinity of $\Rbr$, defined as the radius at which $\tau\sim c/v_{\rm sh}(r)$. Next, note that the formation of the collisionless shock is not restricted to the assumed density profile, $\rho\propto r^{-2}$. Consider the more general case of an ejecta with mass $M_{\rm ej}$ and velocity $v_{\rm ej}$ propagating into a dense CSM with (accumulated) mass profile $M(r)$. The shock propagates with velocity $v_{\rm sh}\sim v_{\rm ej}$ as long as $M_{\rm ej}\gtrsim M(r)$. Once the shock reaches a point where the optical depth is less than $c/v_{\rm ej}$ breakout occurs. The considerations above imply that CSM at radii $r\gg  \Rbr$ for which $M(r)\ll M_{\rm ej}$ will have to be accelerated to $v\sim v_{\rm ej}$ by a colliionless shock.

\paragraph*{Emission from thermal electrons.}
The collisionless shock heats the protons on a time scale of $\om_p^{-1}$ to a temperature roughly given by
\begin{equation}
T_p\sim \frac{3}{16}m_pv^2\sim 0.4v_9^2m_ec^2\sim 0.2v_9^2\MeV.
\end{equation}
The electron temperature depends on the unknown amount of collisionless heating. A lower limit for the electron temperature can be obtained by assuming that there is no collisionless heating.

The collisional heating rate of the electrons due to Coulomb collisions with the protons is given by
\begin{eqnarray}
-\frac{dT_p}{dt}=\frac{dT_e}{dt}&\approx&\lm_{ep}\sqrt{\frac2\pi}\frac{m_e}{m_p}T_p(\frac{T_e}{m_ec^2})^{-3/2}n_d\sig_Tc,
\end{eqnarray}
where $\lm_{ep}=30\lm_{ep,1.5}$ is the Coulomb logarithm and it was assumed that $T_e/(m_ec^2)\gg T_p/(m_pc^2)$.
The fastest possible cooling source for thermal electrons is Inverse Compton (IC) scattering of the local radiation field, which carries a significant fraction $\ep_{\gamma}\lesssim1$ of the post shock energy and is given by
\begin{equation}
\frac{dT_e}{dt}=-\frac{2}{3}\frac{4T_e}{m_e c^2}\sig_TcU_{\gamma},
\end{equation}
where $U_{\gamma}=\ep_{\gamma}n_dT_d$ is the photon energy density and $n_d$ is the shocked material proton density.
Assuming $U_{\gamma}\lesssim n_dT_p$ (equivalently, $\ep_{\gamma}\lesssim 1$) we find
\begin{eqnarray}\label{eq:minhnu}
\frac{T_e}{m_ec^2}\gtrsim 0.6\left(\ep_{\gamma}\frac{m_e}{m_p}\lm_e\right)^{2/5}
\soo T_e\gtrsim 60 \keV \lm_{ep,1.5}^{2/5}.
\end{eqnarray}

The time it takes the protons to lose a significant fraction of their energy (which is of the order of the total available energy) is
\begin{eqnarray}
t_{\rm p}&=&T_p\left(\frac{dT_p}{dt}\right)^{-1}\lesssim0.6\left(\lm_{ep}\frac{m_e}{m_p}\right)^{-2/5}\ep_{\gamma}^{3/5}(n_d\sig_Tc)^{-1}\cr
&\sim&3\lm_{ep,1.5}^{-2.5}\ep_{\gamma}^{3/5}(n_d\sig_Tc)^{-1}.
\end{eqnarray}
The proton cooling time is thus much shorter than the dynamical time $\Rbr/v= (c/v)^2/(n\sig_Tc)$,  where $n=\rho/m_p$ is the proton number density in the pre-shocked region and is smaller than $n_d$ by the compression factor. This is not surprising. While the shock is radiation mediated, radiation energy equal to the mechanical energy is generated on each shock crossing time scale. At breakout, the shock crossing time scale equals the dynamical scale and radiation with energy density comparable to the total energy density must be generated during the dynamical time scale. In fact, since the electron temperature is higher than that expected in a corresponding radiation mediated shock, the emission efficiency is even higher.

The shock is strongly radiative and the energy is efficiently converted to radiation. The typical photon energies are expected to be of the same order of magnitude as the electron energies, i.e. $\gtrsim 60\keV$.
The calculation of the emitted spectrum is beyond the scope of this paper.
We note that since the initial photon energies are much lower ($\sim1$~eV assuming equilibrium) we expect that the spectrum hardens continuously with time and that on the breakout time scale, significant emission is likely emitted at all intermediate energies.

We conclude that gamma-rays/hard X-rays will be emitted with total energy comparable to that of the breakout energy
\begin{eqnarray}\label{eq:Th_gamma}
E_{\gamma}=\frac{4\pi \Rbr^2}{\sig_T}m_pcv\sim 10^{49} v_9R_{14}^2\erg
\end{eqnarray}
on a time scale similar to the breakout time
$t\sim R/v\sim 1R_{14}v_9^{-1}\dy$ with typical luminosity
\begin{eqnarray}
L_{\gamma}\sim 10^{45} R_{14}v_9^2\erg \sec^{-1}.
\end{eqnarray}

When the shock expands, it will remain radiative beyond $\Rbr$ and the total emitted energy, integrated over longer time scales, may be significantly larger than that of the breakout energy.

\paragraph*{Accelerated protons: Non-thermal emission energy.}
Relativistic particles (CRs) accelerated in the collisionless shock that forms due to the collision of the SN ejecta with dense interstellar material may emit high energy gamma rays and neutrinos due to the interaction with the dense material \cite{Murase10}. The collisionless shock that was shown above to be produced during breakout from a dense wind is a constrained example of such interaction and may be a source of detectable high energy neutrinos and gamma rays.
Here we focus on the emission from accelerated protons and their products. In what follows it is assumed that the accelerated protons carry a fraction $\ep_{CR}=0.1\ep_{CR,-1}$ of the post shock energy and have a flat power law energy distribution, $\vep^2dn/d_{\vep}\propto \vep^{0}$.

The cooling time of a relativistic accelerated proton due to inelastic pp collisions is roughly given by
\begin{eqnarray}
t_{pp}&=&(0.2(\rho/m_p)\sig_{pp}c)^{-1}=5\frac{\sig_T}{\sig_{pp}}\left(\frac{v}{c}\right)^2\frac{\Rbr}{v}\cr
&\sim&0.1v_9^{2}\frac{\Rbr}{v}.
\end{eqnarray}
Hence, for slow enough shock velocities, $v/c\lesssim 0.1$, protons accelerated at breakout efficiently convert their energy to neutrinos, gamma-rays and pairs by pion production and decay (and muon decay). In this section we restrict the discussion to $v/c\lesssim 0.1$. For such shock velocities, the amount of energy emitted by relativistic protons during breakout is expected to be roughly a fraction $\ep_{CR}$ of the energy emitted by the thermal particles. Using Eq.~\eqref{eq:Th_gamma} we have
$
E_{\rm Non-thermal}\sim 10^{48}\ep_{CR,-1}R_{14}^2v_9\erg.
$

At later stages, $t_{pp}v/r$ grows linearly with $r$, and as long as it is smaller than unity, the energy converted into pions increases linearly with the accumulated mass. The radius $r_{pp}$ at which the proton energy loss time is equal to the dynamical time, $t_{pp}v/r_{pp}=1$, is
\begin{equation}\label{eq:TenR}
r_{pp}\sim10 v_9^{-2}\Rbr.
\end{equation}
Beyond this radius, the fraction of energy converted to pions drops like $1/r$ ($t_{pp}\propto \rho^{-1}\propto r^{2}$ while the available energy increases linearly with $r$) implying a logarithmic increase in the total emitted energy.  Given that in reality, $v(r)$ is slowly declining, the total contribution to the non thermal fluence from $r>r_{pp}$ is of order unity compared to fluence produce up to this radius. The total emitted energy is therefore given by
\begin{equation}\label{eq:E_NonTh}
E_{\rm Non-thermal}\sim 10^{49}\ep_{\rm CR,-1}R_{14}^2v_9^{-1}\erg.
\end{equation}

\paragraph*{Accelerated protons: Maximal proton energy.}
The maximal proton energy is limited by the time available for acceleration which is the shorter of the dynamical time and energy loss time. The acceleration time depends on the unknown magnetic field value and the loss time depends on the unknown target photon energy distribution. Nevertheless, we next demonstrate that protons are very likely to be accelerated to at least multi-TeV energies.

Assuming Bohm diffusion, the acceleration time to energy $\vep$ is given by
 \begin{equation}\label{eq:tacc}
t_{\rm acc}=\frac{\vep}{(v/c)^2eBc}\sim 2\times 10^{-7} \frac{\vep_{\rm TeV}}{\ep_B^{1/2}v_9^{3/2}R_{14}^{1/2}}\frac{\Rbr}{v},
\end{equation}
where $B$ is the post shock magnetic field and $\ep_B=B^2/(8\pi\rho v^2)$ is roughly the fraction of postshock energy carried by it.
For $\TeV$ CRs, the acceleration time is thus much shorter than the dynamical time and the $pp$ energy loss time. For protons in the range $10-1000\TeV$ the strongest possible cooling mechanism is photo-production of pions, with cooling time
\begin{eqnarray}\label{eq:photoproduction}
t_{p\gamma}=(0.2n_{\gamma}\sig_{p\gamma}c)^{-1}\gtrsim5\frac{\sig_T}{\sig_{p\gamma}}\frac{h\nu_{\gamma}}{m_pc^2}\frac{\Rbr}{v},
\end{eqnarray}
where $n_{\gamma} (h\nu_{\gamma})$ is the target photon number density (typical energy) and  we conservatively assumed that $n_{\gamma}=\rho v^2/(h\nu_{\gamma})$. Photo-production of pions occurs if the proton energy is higher than the threshold, $\sim m_{\pi}m_pc^4/(h\nu_{\gamma})\sim 0.13(h\nu_{\gamma}/\MeV)^{-1} \TeV$. The possible presence of many $\sim 1\MeV$ photons implies that photo-production may be important for $1\TeV$ protons. Photo production is not important if the target photons have $\sim 1\eV$ energies, as assumed in \cite{Murase10}. Given the constraint $n_{\gamma}\lesssim\rho v^2/(h\nu_{\gamma})$, the strongest losses for protons of energy $\vep=\vep_{\rm TeV}\TeV$ occurs for target photons having a typical energy of $h\nu_{\gamma}\sim m_{\pi}m_pc^4/\vep\sim 0.13 \MeV \vep_{\rm TeV}^{-1}$. Using this in Eq. \eqref{eq:photoproduction} we obtain
\begin{eqnarray}\label{eq:photoproduction2}
t_{p\gamma}&\gtrsim&1\vep_{\rm TeV}^{-1}\frac{\Rbr}{v}.
\end{eqnarray}
Comparing Eq. \eqref{eq:photoproduction2} to Eq. \eqref{eq:tacc} we conclude that acceleration to multi TeV energies is possible for $\ep_B\gtrsim 10^{-13}v_9^{-3}R_{14}^{-1}\vep_{\rm TeV}^4$, implying that reaching energies well above $1~\TeV$ is very likely.

We verified that proton CRs do not suffer significant losses due to Inverse Compton and Synchrotron emission during the acceleration time, and that the resulting pions and muons do not suffer significant energy losses due to these processes before decaying. Finally, note that the maximal proton energy is increasing with radius since the ratio of proton acceleration time to dynamical time is independent of radius ({$t_{\rm acc}\propto B^{-1}\propto \rho^{-1/2}\propto r$}) while the ratio of all the loss times to the dynamical time decreases with radius.

\paragraph*{Accelerated protons: Multi TeV neutrinos.}
Roughly a third of the non thermal energy Eq. \eqref{eq:E_NonTh} will be emitted in muon neutrinos (and anti-neutrinos) and a significant fraction of this energy may be emitted beyond TeV energies.
In the neutrino energy range of one to hundred TeV, the effective area for muon neutrinos of a Cherenkov neutrino detector like IceCube is increasing linearly with energy, approximately as $10^{-6}\vep_{\rm TeV}A$, where $A=10^{10}A_{10}\cm^2$ is the geometrical cross-section of the detector. The number of muons induced by one to hundred TeV neutrinos  is therefore independent of the neutrino spectrum in this range and is given by
\begin{equation}\label{eq:ObsNeutrinos}
N_{\mu}\sim 5 \frac{E_{\nu_\mu,1-100\rm TeV}/10^{51}\erg}{(d/100\Mpc)^2}\sim 1\frac{\ep_{p,-1}R_{15}^2}{v_9 (d/100\Mpc)^2}
\end{equation}
where $d$ is the distance to the SN and where we optimistically assumed that $1/3$ of the non thermal emission, Eq. \eqref{eq:E_NonTh}, is in multi TeV neutrinos.

\paragraph*{Accelerated protons: Gamma rays.}
High energy gamma-rays and pairs with energies reaching multi TeV energy will be generated with a comparable rate to that of the neutrinos. The pairs will emit further high energy gamma- rays by Inverse Compton interactions with the radiation field. Emission below $\sim1~\MeV$ will be mixed with the emission from the thermal electrons. Emission at a photon energy $h\nu\gtrsim \MeV$ may be suppressed by the large optical depth for pair creation, which depends on the density of photons with energies above the pair production threshold $h\nu_T\gtrsim m_ec^2/(h\nu)$.

An upper limit to the optical depth for pair creation at a given photon energy, $h\nu$, can be obtained by using the fact that the total energy density of photons of any frequency is smaller than $\rho v^2$. Assuming that the energy of photons per unit logarithmic frequency does not exceed $\ep_{0.1}\times10\%$ of $\rho v^2$, and focusing on the radius $10\Rbr$ at which the protons are still efficiently cooled (for $v\sim 10^9\cm\sec^{-1}$, see Eq. \eqref{eq:TenR}), we find
\begin{equation}
\tau_{\gamma \gamma}\lesssim \ep\frac{v}{c}\frac{m_p}{m_e}\frac{h\nu}{m_ec^2}\sim\ep_{0.1} 0.6 v_9 \frac{h\nu}{m_ec^2}.
\end{equation}
The emitted spectrum is suppressed by at most $\sim \tau^{-1}$. In this 'worst case scenario', such bursts will be too faint to be observable by high energy ($h\nu\sim1$~GeV) gamma-ray detectors such as Fermi.

\paragraph*{Increased optical depth due to electron-positron pairs.}
A significant fraction of the energy carried by the accelerated protons may be converted to electron positron pairs. The number of pairs per proton is limited by the available energy to  $n_{e\pm}/n\lesssim \ep_{CR} m_pv^2/(m_e c^2)\sim 0.2\ep_{CR,-1} v_9^2$. The presence of pairs cannot change the optical depth significantly for $v/c\lesssim 0.1$. For very fast shocks, $v\sim c$, the presence of pairs can potentially increase the optical depth considerably and affect the emitted radiation.

\paragraph*{Discussion.}
We have shown that shock breakouts in optically thick winds will necessarily be accompanied by high energy radiation from a collisionless shock that  inevitably forms on the time scale of the breakout outburst.

Low luminosity GRBs associated with SNe have been suggested to be the outbursts associated with fast shocks $v\gtrsim 0.1c$  breaking out of dense optically thick winds \cite{Campana06,Waxman07,Wang07,Soderberg08,Katz10}.
As we have shown here, a significant fraction of the observed radiation, or even most of it, may be generated by the collisionless shock that will form during the breakout.

If the slow, $v/c\sim0.03$, breakout interpretation of events such as PTF09u \cite{Ofek10} is correct, a significant amount of energy, $E\sim 10^{51}\erg$, is expected to be emitted in hard X/$\gamma$-rays reaching energies $h\nu\gtrsim 50\keV$, Eq.~\eqref{eq:minhnu}, and multi-TeV neutrinos \cite[see also][]{Murase10}. X-rays at lower energies are likely to be emitted with similar efficiency and would be easily detected by instruments like the X-ray telescope (XRT) on board Swift or the Chandra X-ray observatory. TeV neutrinos may be detectable by experiments like IceCube, see Eq. \eqref{eq:ObsNeutrinos}, provided such events are sufficiently common and a similar event occurs at a distance $d\lesssim 100 \Mpc$ (compared to $\sim 300 \Mpc$ for PTF09uj).

Finally, we note that if the CSM breakout explanation of very luminous SNe \cite[e.g.][]{Quimby07} is correct, our analysis implies that these events should be accompanied by strong high energy X-ray emission. Lack off \cite{Miller09}, or very weak \cite{Smith07a}, X-ray emission from some of these events challenges this interpretation.

B.K. is supported by NASA through Einstein Postdoctoral Fellowship awarded by the Chandra X-ray Center, which is operated by the Smithsonian Astrophysical Observatory for NASA under contract NAS8-03060.
The research of E.W and N.S is partially supported by ISF, Minerva and PBC grants.

\end{document}